\documentclass[conference]{IEEEtran}

\usepackage{cite}
\usepackage{graphicx}
\DeclareGraphicsExtensions{.eps,.png}

\usepackage[cmex10]{amsmath}
\usepackage[caption=false,font=footnotesize]{subfig}
\usepackage{amssymb}
\usepackage{amsmath}
\usepackage{subeqnarray}
\usepackage{cases}%
\usepackage{float}
\usepackage{multirow}

\begin{document}

\bibliographystyle{ieeetr}

\title{Two-tier Spatial Modeling of Base Stations in Cellular Networks}

\author{\IEEEauthorblockN{Yifan Zhou\IEEEauthorrefmark{1}\IEEEauthorrefmark{2},
Zhifeng Zhao\IEEEauthorrefmark{1}\IEEEauthorrefmark{2}, Qianlan Ying\IEEEauthorrefmark{1}\IEEEauthorrefmark{2}, Rongpeng Li\IEEEauthorrefmark{1}\IEEEauthorrefmark{2}, Xuan Zhou\IEEEauthorrefmark{1}\IEEEauthorrefmark{2}, and
Honggang Zhang\IEEEauthorrefmark{1}\IEEEauthorrefmark{2}\IEEEauthorrefmark{3}}
\IEEEauthorblockA{\IEEEauthorrefmark{1}York-Zhejiang Lab for Cognitive Radio and Green Communications}
\IEEEauthorblockA{\IEEEauthorrefmark{2}Dept. of Information Science and Electronic Engineering\\Zhejiang University, Zheda Road 38, Hangzhou 310027, China\\Email: \{zhouyftt, zhaozf, greenjelly, lirongpeng, zhouxuan, honggangzhang\}@zju.edu.cn}
\IEEEauthorblockA{\IEEEauthorrefmark{3}Universit\'{e} Europ\'{e}enne de Bretagne \& Sup\'{e}lec, Avenue de la Boulaie, CS 47601, 35576 Cesson-S\'{e}vign\'{e}  Cedex, France}}
\maketitle

\begin{abstract}
Poisson Point Process (PPP) has been widely adopted as an efficient model for the spatial distribution of base stations (BSs) in cellular networks. However, real BSs deployment are rarely completely random, due to environmental impact on actual site planning. Particularly,  for multi-tier heterogeneous cellular networks, operators have to place different BSs according to local coverage and capacity requirement, and the diversity of BSs' functions may result in different spatial patterns on each networking tier. In this paper, we consider a two-tier scenario that consists of macrocell and microcell BSs in cellular networks. By analyzing these two tiers separately and applying both classical statistics and network performance as evaluation metrics, we obtain accurate spatial model of BSs deployment for each tier. Basically, we verify the inaccuracy of using PPP in BS locations modeling for either macrocells or microcells. Specifically, we find that the first tier with macrocell BSs is dispersed and can be precisely modelled by Strauss point process, while Matern cluster process captures the second tier's aggregation nature very well. These statistical models coincide with the inherent properties of macrocell and microcell BSs respectively, thus providing a new perspective in understanding the relationship between spatial structure and operational functions of BSs.
\end{abstract}
\IEEEpeerreviewmaketitle

\section{Introduction}
Network topology has a great impact on the performance of cellular networks, since the power of received signal varies depending on the distance between transmitter and receiver. Moreover, interference characterization become even more complicated due to path loss and multipath fading effect\cite{andrews2010primer}. Hence, realistic spatial modeling of base stations (BSs) is essential for accurate performance evaluation in cellular networks.

In recent years, Poisson point process (PPP) has been proposed as an efficient way to model wireless network structures \cite{haenggi2009stochastic}. The PPP model for characterizing BS locations can provide tractable and useful results for key performance evaluation in cellular networks\cite{andrews2011tractable}. However, it may not be a practically solid model for BSs distribution because its complete randomness is unrealistic in real deployment \cite{riihijarvi2010modeling,taylor2012pairwise,guo2013spatial,lee2013stochastic,deng2014ginibre,wu2014spatial}.

Indeed, BSs are not independently distributed in certain areas for purpose of optimizing coverage and capacity performance. In \cite{riihijarvi2010modeling}, the authors discover that the Geyer saturated process, which takes account of pairwise interaction between BSs, can accurately reproduce the spatial structure for various wireless networks. More specifically in cellular networks, Geyer saturation process and its special case Strauss process are utilized to model macrocell deployment for different scenarios in \cite{taylor2012pairwise}. Besides, Poisson hard-core process (PHCP) is also proposed to model BSs locations in \cite{guo2013spatial}, while it is still shown that Strauss process provides the best fit in terms of coverage probability. Poisson cluster process is verified to be able to model urban area deployment in \cite{lee2013stochastic}. Very recently, the Ginibre point process has been proposed as a suitable model for wireless networks with nodes repulsion\cite{deng2014ginibre}, in the light of compromise between accuracy and tractability. All these studies above put forward corresponding spatial models for BSs deployment, but the types of BSs (i.e. macrocell or microcell) have not been taken into account seriously.

As we know, cellular networks have been undergoing an evolution towards heterogeneous networking architecture. System performance evaluation and resource allocation become more complicated for multi-tier scenario, since each tier may differ in transmit power, coverage area and supported rate\cite{dhillon2012modeling}\cite{cheung2012throughput}. This lasting trend highlights the importance for hierarchical spatial analysis of multi-tier structure. In the typical case of two-tier cellular networks, the spatial structures of macrocell and microcell deployment have not been analyzed separately ever, despite of their inherent differences in functionality and networking feature.

Actually, macrocell BSs are neither too close nor too far away from each other in order to satisfy coverage requirement and decrease inter-cell interference. Therefore, there is repulsion between macrocell BSs. On the other hand, microcells are usually deployed to diminish coverage hole and offload network traffic, which always exhibits aggregation feature. So microcell BSs would be clustered. These facts provide reasonable basis to adopt Gibbs and Neyman-Scott processes\cite{chiu2013stochastic} for modeling macrocell and microcell deployment, respectively.

This paper aims to find realistic spatial models for macrocell and microcell BSs separately. Based on massive and detailed real data from one of the largest operators in China, we obtain enough location records for fitting and testing models. By applying statistical metrics in point process field such as $L$-function and nearest neighbor distance distribution\cite{ripley1991statistical}, along with performance metrics in cellular network analysis such as coverage probability and cell coverage area, different hypotheses of BSs deployment on each tier are tested. Accordingly, the main contributions of this paper are summarized as follows:
\subsubsection{\bf{Verification of the inaccuracy of PPP}}
Different with the related works, the validity of PPP for BS locations on different tier is tested and we verify its inaccuracy in terms of classical spatial statistical metric for both macrocells and microcells.
\subsubsection{\bf{Realistic spatial model for macrocells}}
We find that macrocell BSs are regularly deployed and can be well modeled by Strauss point process. The fitting result coincides with the operational function of macrocells which is targeted at providing overall coverage in cellular networks.
\subsubsection{\bf{Realistic spatial model for microcells}}
Different with macrocells, we discover that microcell BSs tend to be more clustered, and can be accurately reproduced by Matern cluster process (MCP), which is consistent with microcells' main functionality of diminishing coverage hole and traffic offloading.

\section{Background-Point Process Theory}
Gibbs model\cite{chiu2013stochastic} is a kind of general statistical model that can be characterized by probability density, which is helpful in fitting and simulation using Monte Carlo method. Without loss of generality, we consider a point pattern $\mathbf{z} =\lbrace z_1,z_2,...,z_{n(\mathbf{z})}\rbrace $ placed in a bounded window $W$, where $n(\mathbf{z})$ is the number of points in $\mathbf{z}$. For simplicity, only pairwise interaction is considered here, and its probability density function (PDF) can be defined as:
\begin{equation}
       f(\mathbf{z})= \alpha\cdot[\prod_{i=1}^{n(\mathbf{z})}\mu(z_i)]\cdot[\prod_{i<j}\rho(z_i,z_j)],
\end{equation}
where $\alpha$ is a normalizing factor to ensure the integral to unity, $\mu(z_i)$ are functions modeling the first order property, and $\rho(z_i,z_j)$ are functions modeling pairwise interaction. By setting $\mu(z)$ a constant $\beta$, and defining $\rho(z_i,z_j)$ as follows:
\begin{equation}
\rho(z_i,z_j)=\left\{
\begin{array}{ll}
1, & \parallel z_i-z_j \parallel > r\\
\gamma, & \parallel z_i-z_j\parallel \leq r
\end{array} \right.,
\end{equation}
the PDF is simplified to
\begin{equation}
f(\mathbf{z})=\alpha\beta^{n(\mathbf{z})}\gamma^{p(\mathbf{z})},
\end{equation}
where $p(\mathbf{z})$ is the number of point pairs that are less than $r$ units apart. If $\gamma=1$, there is no interaction between points, and it can be simplified to PPP. By adding a constraint that no distinct points are allowed to come closer than distance $hc$, the PPP further reduces to Poisson hardcore process (PHCP) with hard core distance $hc$. Clearly, $\gamma<1$ indicates the repulsive case, and the density decreases with $p(\mathbf{z})$, Strauss point process is one of the representative examples. Otherwise, when $\gamma>1$, in order to make the PDF integrable and capable of modeling clustering effect, a saturation threshold is added in the exponent and the PDF is:
\begin{equation}
f(\mathbf{z})=\alpha\beta^{n(\mathbf{z})}\gamma^{\min(p(\mathbf{z}),sat)}.
\end{equation}
It's termed as Geyer saturation process, a generalization of Strauss point process. Moreover, it would reduce to a PPP for $sat=0$, or a Strauss point process for $sat\to\infty$.

The Neyman-Scott process\cite{chiu2013stochastic} consists of the set of clusters of offspring points, centered around an unobserved set of parent points. Matern cluster process (MCP) is a special case of the Neyman-Scott process, where the number of offspring points per cluster is Poisson distributed with intensity $\lambda_c$, and their positions are placed uniformly inside a disc of radius $R$ centred on the parent points. We assume that the cluster centers form the point pattern $\mathbf{c}$ which is Poisson distributed with intensity $\lambda_p>0$, while conditional on $\mathbf{c}=\{c_1,c_2,...,c_n\}$ associate each $c_i$ with a Poisson point process $\mathbf{x}_i$ with intensity $\lambda_c>0$ and these offspring point processes are independent with each other. The density function of Matern cluster process can be written as:
\begin{equation} \
f(\xi-c_i)=\frac{2r}{R^2}, \qquad \text{for} \quad r=\parallel\xi-c_i\parallel \leq R.
\end{equation}
All these models mentioned above can be fitted using maximum pseudolikelihood method\cite{baddeley2000practical} provided in $Spatstat$, an R package\cite{baddeley2005spatstat}.

\section{Evaluation Statistics and Fitting Method}
In order to obtain the separate spatial models of two-tier BSs, we investigate a dense urban area in a prosperous city in China. BSs of GSM cellular networks are considered here only for representativeness and consistency. The selected $3\times3$ $km^2$ square area contains 266 BSs including 77 macrocells and 189 microcells, whereas the high BS density implies strong coverage and capacity demands. The macrocell BSs are referred to as point pattern $\mathbf{x}$, while point pattern $\mathbf{y}$ is for microcell BSs. Both of them have been mapped by the same scale onto a unit square as seen in Fig.1.

\begin{figure}[!t]
\centering
\includegraphics[trim=0mm 10mm 0mm 10mm,clip,width=0.4\textwidth]{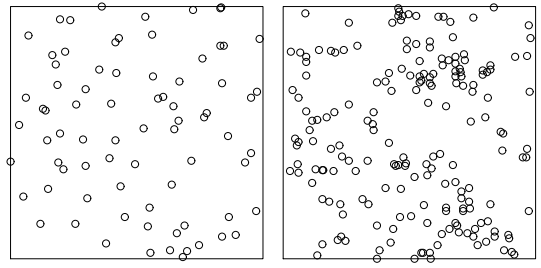}
\caption{($left$): Macrocells as point pattern $\mathbf{x}$ exhibits inhibition. ($right$): Microcells in the same area as point pattern $\mathbf{y}$ appears to be clustered.}
\end{figure}

\begin{figure}[!t]
\centering
\includegraphics[trim=0mm 5mm 0mm 15mm,clip,width=0.35\textwidth]{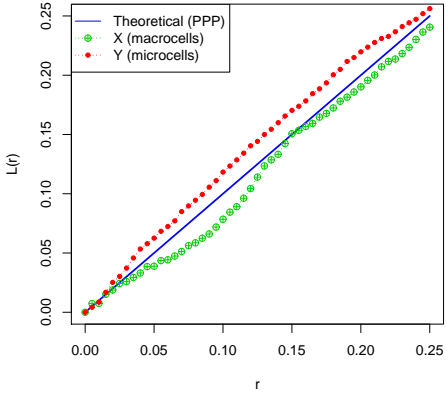}
\caption{$L$-function of point pattern $\mathbf{x}$ and $\mathbf{y}$, compared with the theoretical curve for PPP.}
\end{figure}

For hypothesis testing, we use the following four evaluation statistics including two classical metrics and two network performance metrics:
\setcounter{subsubsection}{0}
\subsubsection{\bf{L-function}}
$L$-function is a transformation of the Ripley's $K$-function ($K(r)$), which is widely used to test the validity of a point process in stochastic geometry\cite{ripley1991statistical}. It reflects regularity or clustering property of a point pattern and is defined as:
\begin{equation}
L(r)=\sqrt{\frac{K(r)}{\pi}}.
\end{equation}
For a completely random (uniform Poisson) point pattern, the theoretical value is $L(r)=r$, which is used as a baseline to judge a point pattern's spatial character\cite{ripley1991statistical}. If $L(r)<r$, then there is dispersion on this $r$ scale and should be modeled by an repulsive process; otherwise it is aggregated if $L(r)>r$ and should be modeled by a clustering point process. Due to its explicitness and importance, $L$-function is used as the first-step metric in this paper.

\subsubsection{\bf{Nearest Neighbor Distance Distribution}}
Nearest neighbor distance distribution function of a point process $\mathbf{z}$ is the cumulative distribution function (CDF) $G(r)$ of the distance from a typical random point of $\mathbf{z}$ to the nearest other points of $\mathbf{z}$\cite{ripley1991statistical}. The estimate of $G(r)$ is a useful statistic summarising the clustering property of point pattern by comparing with the theoretical $G(r)$ of a PPP which is
\begin{equation}
G(r)=1-\exp(-\pi\lambda r^2).
\end{equation}
This statistic is utilized as a useful evaluation metric in microcells spatial modeling, since it represents the internal structure of clusters in real BSs deployment.

\subsubsection{\bf{SIR Distribution}}
In order to find a realistic model, we choose signal-to-interference-ratio (SIR) as an evaluation metric to bridge the modeling validity and actual network performance. Assuming each mobile user connects to the BS at location y that offers the highest received SIR. The resulting SIR in position $z$ is defined as:
\begin{equation}
SIR(s, \mathbf{z})= \frac{P_yh_yd(s,y)^{-\alpha}}{\sum_{x\in\mathbf{z}\backslash y}P_xh_xd(s,x)^{-\alpha}},
\end{equation}
where Rayleigh fading is adopted as $h_x$, $h_y\sim\exp(1)$, and the path loss exponent $\alpha$ is assumed to be 4 considering dense urban scenario. $P_x$ and $P_y$ are transmit powers of the corresponding BSs.

\subsubsection{\bf{Voronoi Cell Area Distribution}}
The Voronoi cell of a node $z\in\mathbf{z}$ is defined as $\{y\in\mathbf{R}^2:d(y,x)>d(y,z),\forall x\in\mathbf{z}\backslash z\}$. It is proposed as an evaluation metric due to its similarity with the coverage region of BSs\cite{taylor2012pairwise}, which is a valuable parameter in practical network operations. As mentioned in Section I, macrocell BSs are responsible for coverage-centric requirement, therefore we use this metric as an evaluation statistic in the hypothesis testing of point pattern $\mathbf{x}$.

\section{Proposed Models and Fitting Results}
In order to obtain the accurate point process model for each tier, we first fit the candidate models to the BSs data set using maximum pseudolikelihood method and get the corresponding parameters. To test models' validity, we generate 600 realisations for each fitted model, and for each realization the evaluation statistics are computed to obtain the simulation envelope. After that, we throw out the 30 highest and 30 lowest values to create 90$\%$ confidence intervals for judgement.

\begin{table}[h]
\centering
\caption{Parameters of fitted models.}
\begin{tabular}{cll}
\hline
Point Pattern & Fitted Model & Parameters \\
\hline
                              &   PPP   & $\lambda=20.306$            \\
       $\mathbf{x}$           &  PHCP   & $h_c=0.0047$       \\
        (macrocells)          & Strauss & $r=0.085, \gamma=0.3547$         \\
                              &  Geyer  & $r=0.12, sat=3, \gamma=0.2448$  \\
\hline
       $\mathbf{y}$           &   PPP   & $\lambda=37.837$             \\
        (microcells)          &  Geyer  & $r=0.05, sat=5, \gamma=1.4011$ \\
                              &   MCP   & $\lambda_p=71.552, \lambda_c=2.641, R=0.087$ \\
\hline
\end{tabular}
\end{table}

\subsection{Macrocell Point Pattern Modeling}

The macrocell point pattern $\mathbf{x}$ consists of 77 points. Firstly, we use $L$-function to determine whether the point pattern is repulsive or clustered. As shown in Fig.2, the $L(r)$ of $\mathbf{x}$ is under the $L(r)=r$ (PPP) baseline, which indicates that macrocell BSs tend to be dispersively distributed. Secondly, given the repulsive property, we adopt the Strauss process, PHCP, and Geyer saturation process for the following modeling and use the PPP as a benchmark for comparison. Applying maximum pseudolikelihood method\cite{baddeley2000practical}, we obtain these fitted models and their corresponding parameters in Table I. In order to test the validity of each model, we utilize the confidence interval of evaluation statistics for further verification.

Explicitly, the $L$-function of $\mathbf{x}$ is presented in Fig.3 along with the fitted PPP and Geyer process envelope, while the PHCP and Strauss envelope are depicted in Fig.4. We observe that the real $L(r)$ is mostly below the PPP envelope, especially in the range of $0.05<r<0.15$ which verifies the irrationality of PPP's complete randomness. Similarly, Geyer process is unable to capture the characteristics of $\mathbf{x}$ since its $L$-function envelope can't always surround that of the data set. In Fig.4, the hypothesis that $\mathbf{x}$ is PHCP can also be rejected, but Strauss process gives a perfect fitting according to the confidence interval. Conclusively, we can reject the PPP, PHCP and Geyer process hypothesis by the $L$-function statistic evaluation, and the Strauss process is shown to be a suitable model for macrocell deployment. Furthermore, Voronoi cell area and coverage probability distributions are applied to reinforce this conclusion, and the related results are illustrated in Fig.5 and 6. The results show that the Strauss process provides an accurate modeling for $\mathbf{x}$ not only in terms of classical statistic metrics but also in practical network performance metrics. Given these hypothesis testing results above, we can claim that the Strauss point process gives the best fit for macrocell deployment.

\begin{figure}[!t]
\centering
\includegraphics[trim=0mm 5mm 0mm 15mm,clip,width=0.35\textwidth]{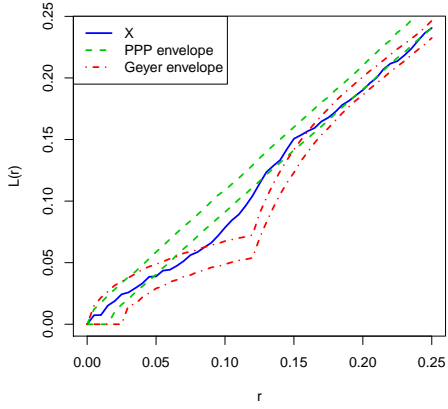}
\caption{Rejection of PPP and Geyer process for $\mathbf{x}$ by $L$-function.}
\end{figure}

\begin{figure}[!t]
\centering
\includegraphics[trim=0mm 5mm 0mm 15mm,clip,width=0.35\textwidth]{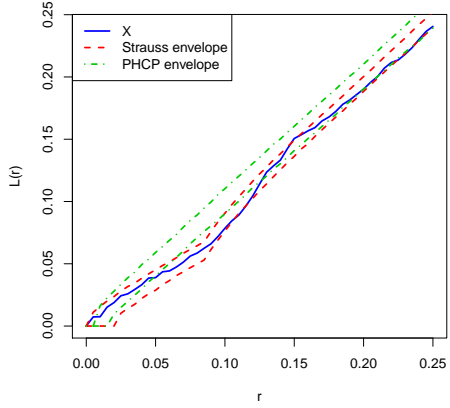}
\caption{$\mathbf{x}$'s $L$-function lies within the fitted Strauss envelope but rejects PHCP.}
\end{figure}

\begin{figure}[!t]
\centering
\includegraphics[trim=0mm 5mm 0mm 15mm,clip,width=0.35\textwidth]{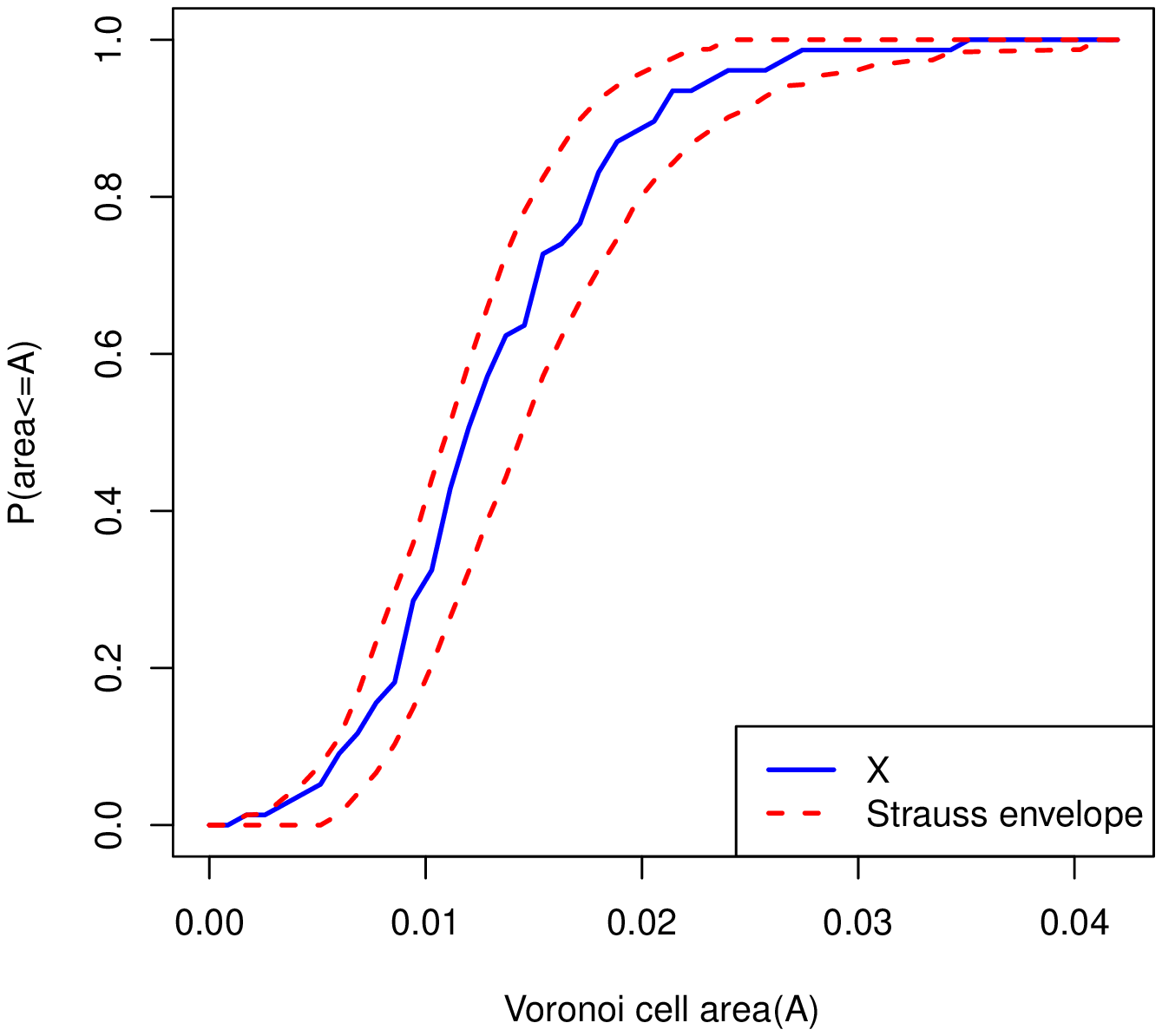}
\caption{$\mathbf{x}$'s Voronoi area distribution lies within the fitted Strauss envelope.}
\end{figure}

\begin{figure}[!t]
\centering
\includegraphics[trim=0mm 5mm 0mm 15mm,clip,width=0.35\textwidth]{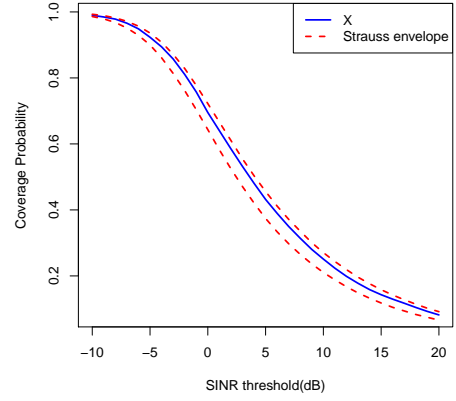}
\caption{$\mathbf{x}$'s SIR distribution lies within the fitted Strauss envelope.}
\end{figure}

\subsection{Microcell Point Pattern Modeling}

The microcell point pattern $\mathbf{y}$ contains 189 points, and thus it has far more points than $\mathbf{x}$ as expected. The $L(r)$ curve of $\mathbf{y}$ in Fig.2 indicates that microcell BSs are aggregately deployed in this area. Hence, in addition to using PPP for inaccuracy verification, the Geyer saturation process and Matern cluster process are considered credible candidates for accurate characterization. The fitted models and corresponding parameters are listed in Table I.

\begin{figure}[!t]
\centering
\includegraphics[trim=0mm 5mm 0mm 15mm,clip,width=0.35\textwidth]{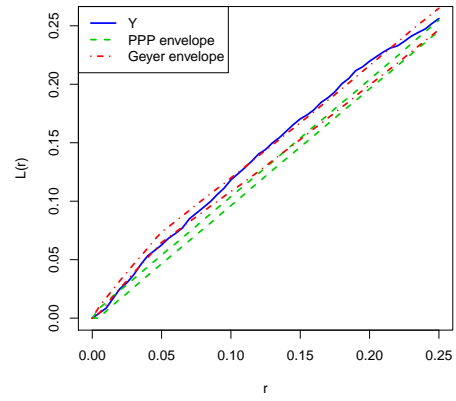}
\caption{Rejection of PPP and Geyer process for $\mathbf{y}$ by $L$-function.}
\end{figure}

\begin{figure}[!t]
\centering
\includegraphics[trim=0mm 5mm 0mm 15mm,clip,width=0.35\textwidth]{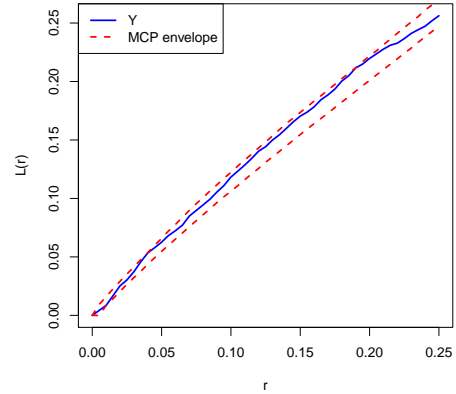}
\caption{$\mathbf{y}$'s $L$-function lies within the fitted Matern envelope.}
\end{figure}

Firstly, we use $L$-function to test each hypothesis. The envelope of PPP and Geyer simulations are presented in Fig.7. As we can see, the $L(r)$ of $\mathbf{y}$ is totally above the PPP's envelope which verifies the inaccuracy of PPP firmly. The same result applies for Geyer process in the range of $0.10<r<0.20$. So we can reject these two hypotheses according to $L(r)$ in the first place. The hypothesis of MCP can be accepted in term of $L$-function metric in Fig.8. In order to sustain this claim, the nearest neighbor distance distribution is applied as an evaluation metric in Fig.9, indicating $\mathbf{y}$'s $G(r)$ lies within the envelope exactly. Moreover, the coverage probability distribution of $\mathbf{y}$ falls into the envelope of fitted MCP in Fig.10, which verifies the MCP's accuracy in terms of network performance metric as well. From these results above, it clearly confirms that MCP can reproduce $\mathbf{y}$'s real deployment accurately, so we conclude that microcell BSs tend to be clustered deployed and can be well modelled by Matern cluster process.

\begin{figure}[!t]
\centering
\includegraphics[trim=0mm 5mm 0mm 15mm,clip,width=0.35\textwidth]{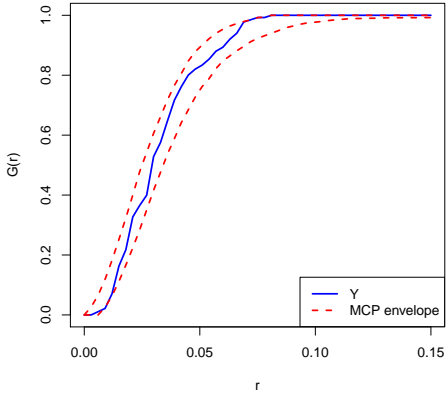}
\caption{$\mathbf{y}$'s $G(r)$ lies within Matern process envelope.}
\end{figure}

\begin{figure}[!t]
\centering
\includegraphics[trim=0mm 5mm 0mm 15mm,clip,width=0.35\textwidth]{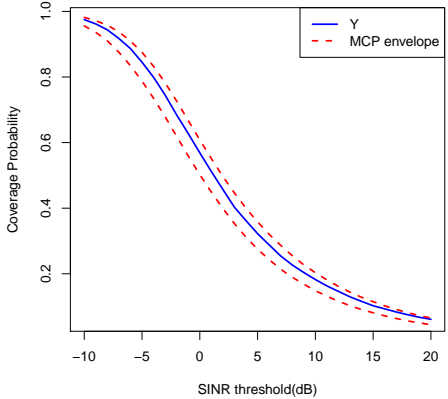}
\caption{$\mathbf{y}$'s SIR distribution lies within the fitted Matern envelope.}
\end{figure}

\section{Conclusions}
In this paper, we proposed a new method for spatial modeling of BS locations and obtained accurate models for both macrocell and microcell deployment in a two-tier cellular network. Specifically, by applying both classical statistics and network performance as fitting metrics, we found that macrocell BSs are dispersedly distributed and can be well modeled by Strauss point process, while microcell BSs present aggregation property and can be accurately reproduced by Matern cluster process.
To the best of our knowledge, it is the first time that different types of BSs (i.e. macrocell or microcell) are spatially modeled separately, hence we provide a new perspective in understanding the fundamental relationship between spatial structure and operational functions of BSs in heterogeneous cellular network. Yet, large-scale verification in different places is still necessary to reinforce this conclusion more generally. Besides, unifying these two separate spatial models in one integrated theoretical expressions can be left as future work, given the potential usefulness for more realistic network performance analysis.


\section*{Acknowledgment}
This paper is partially supported by the National Basic Research Program of China (973Green, No. 2012CB316000), the Key Project of Chinese Ministry of Education (No. 313053), the Key Technologies R\&D Program of China (No. 2012BAH75F01), and the grant of ``Investing for the Future'' program of France ANR to the CominLabs excellence laboratory (ANR-10-LABX-07-01).

\bibliography{C:/Users/zhouyftt/Desktop/writing/paper/shoulders}



%

\end{document}